\documentclass{ws-ijmpd}

\RequirePackage{graphicx}

\usepackage[mathscr]{eucal}
\usepackage[caption=false]{subfig}
\usepackage[super,compress]{cite}
\usepackage{color}

\newcommand{\beq}{\begin{equation}}
\newcommand{\eeq}{\end{equation}}

\newcommand{\bea}{\begin{eqnarray}}
\newcommand{\eea}{\end{eqnarray}}

\newcommand{\beano}{\begin{eqnarray*}}          
\newcommand{\eeano}{\end{eqnarray*}}

\begin{document}

\markboth{Gurzadyan and Kocharyan}
{SUPERSPACE AND STABILITY IN GR}

%
\catchline{}{}{}{}{}
%

\title{SUPERSPACE AND GLOBAL STABILITY\\ IN GENERAL RELATIVITY} 

\author{A.V.~GURZADYAN}

\address{Department of Mathematics and Mathematical Modeling,\\
Russian-Armenian (Slavonic) University,\\ 
Yerevan, Armenia\\
gurzadyana@gmail.com}

\author{A.A.~KOCHARYAN}

\address{School of Physics and Astronomy, Monash University,\\
Clayton, Australia}

\maketitle

\begin{history}
\received{5 December 2016}
\revised{10 December 2016}
\accepted{15 December 2016}
\end{history}

\begin{abstract}
A framework is developed enabling the global analysis of the stability of cosmological models using the local geometric characteristics of the infinite-dimensional superspace, i.e. using the generalised Jacobi equation reformulated for pseudo-Riemannian manifolds. We give a direct formalism for dynamical analysis in the superspace, the requisite equation pertinent for stability analysis of the universe by means of generalized covariant and Fermi derivative is derived. 
Then, the relevant definitions and formulae are retrieved for cosmological models with a scalar field.
\end{abstract}

\ccode{PACS numbers: 98.80.-k, 95.35.+d, 95.36.+x, 96.15.Ef}


\section{Introduction}	

General Relativity (GR) provides a basis for our understanding of the dynamics of the expansion of the universe and
of final phases of stellar evolution. The observational discoveries of the last two decades on the accelerated
expansion of the universe, on the dominance of the dark sector i.e. of the dark energy and dark matter, on the tiny properties
of the cosmic microwave background, etc, have determined the refined consideration of variety of cosmological models within GR. Scalar field cosmological
models are attracting essential interest, since they are not only motivated by the inflationary paradigm regarding the very early evolution of the universe, but scalar field coupled with gravity is associated to the dark energy, to the dark matter in galaxy clusters, see Ref.\refcite{Lin,Peeb,Bar1,Bar2,Bern,Ash,Oi} and references therein.

The stability with respect to the initial conditions and/or perturbations is principal in the theoretical analysis of cosmological models,
justifying or vice versa, diminishing the impact of a given cosmological solution. Methodically GR has its peculiarities, namely, the technique and criteria developed in theory of dynamical systems at least have to be reformulated for Lorentzian (pseudo-Riemannian) manifolds. The structure of Einstein equations has motivated the development of such principal approaches such as the geometrodynamics, ADM formalism, minisuperspace approximations Ref.\refcite{MTW}. Solutions found for minisuperspace models, however, can appear non-typical in superspace, thus outlining the need for the rigorous treatment of stability issues in Wheeler-DeWitt superspace. The revealing of the stability of specific cosmological solutions is among the goals of variety of approaches (see e.g. Ref.\refcite{La,Bar1,Ed} and references therein).   
 
The principal intention of this paper is to define a framework for rigorous formulation of the concept of stability of dynamical systems defined in superspace. Our approach, on the one hand, roots on the ADM formalism Ref.\refcite{ADM}, on the other hand, on the method of geodesic flows and Jacobi equation of the theory of dynamical systems Ref.\refcite{An}. The detailed account of this approach is given in Ref.\refcite{GK1}, while here we concentrate on the reduction of the problem of global stability of cosmological solutions to the one based on local geometric descriptors, and then, we apply the developed formalism for the models with a scalar field.  

\section{Dynamical system formulation of gravity}

\qquad We consider a smooth, $(d+1)$-dimensional differentiable manifold $W$ with a Lorentzian metric $^{(d+1)}g$ that is oriented and time-oriented, representing spacetime. Let $M$ be a smooth, orientable, compact, $d$-dimensional differentiable manifold, and let $i:M \to W$ be an embedding of $M$ into $W$ such that the induced metric on $M$ is Riemannian. Then we may visualize $\Sigma = i(M)$ as a spacelike slice through the spacetime $W$; i.e. a surface of simultaneity for some observer (in some coordinate system).

Now suppose that on some interval $I\in \mathbb{R}$ we define an embedding
$i(t):M \to W$ for each $t\in I$, such that this map is continuous w.r.t. $t$.
Further,
\beano
\tfrac{d}{dt}i(t):M\to TW
\eeano
is the map describing the motion, w.r.t. changes in $t$, of each point of $M$ as it is moved through $W$.

Now let $\mathbf{N}(t)$ be the vector field on $M$ corresponding to the projection of $\frac{di(t)}{dt}$
onto $M$, and let $N(t)$ be the scalar field on $M$ defined as the length 
(w.r.t. $^{(d+1)}g$) of the component of $\frac{di(t)}{dt}$	perpendicular to $M$. 
We may visualize these as follows: if we imagine sliding $M$ through $W$, in some uneven fashion, then $\mathbf{N}(t)$, 
called the ``shift vector field", represents the way we twist, rotate, and stretch $M$ as we slide it, 
while $N(t)$, called the ``lapse function", represents the rate of slide (w.r.t. parameter $t$). 
If $N(t) > 0$ everywhere (so that we always slide $M$ in the same direction w.r.t. ``time" on $W$) then we call $i(t)$ a slicing of $W$.
 
\begin{figure}[!tb]
  \centering
    \includegraphics[width=0.8\textwidth]{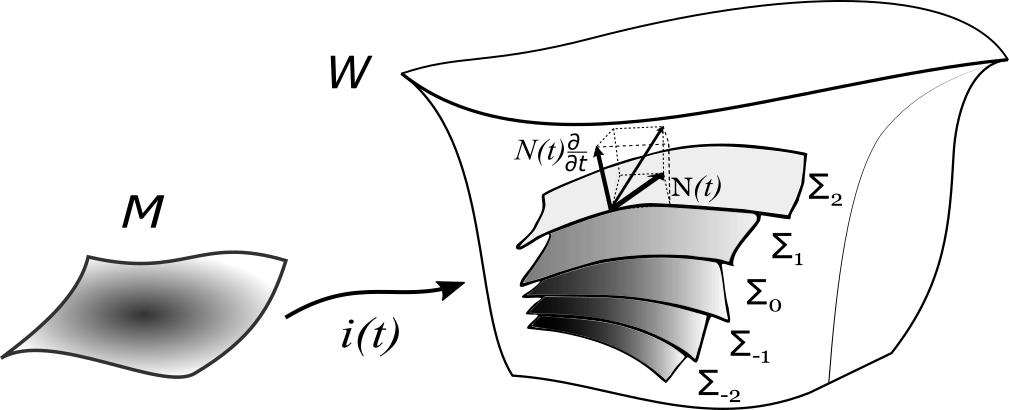}
		  \caption{Embedding of space $M$ into spacetime $W$, thus $W$ is described as a series of space slices and ``shift vectors'' corresponding to sliding of $M$.}
			\label{fig:MtoW}
\end{figure}

Now, if we have a coordinate system on $M$ with coordinates $(x^1, \ldots, x^d)$, 
then we may use $(t, x^1, , \ldots, x^d)$ as a coordinate system on $I\times M$. In this case it is easy to show that the metric on $W$, in these coordinates, is given by:
\beq
^{(d+1)}g_{\alpha\beta}dx^\alpha dx^\beta =-N^2dt^2 +g_{ab}(dx^a+N^adt)(dx^b+N^bdt)\,,
\eeq
where $N^a$ are components of $\mathbf{N}$, and $g_{ab}$ is Riemannian metric on $M$ induced by the embedding $i(t)$.

Thus by means of this slicing we have divided the spacetime into space (manifold $M$) and time (see Fig.\ref{fig:MtoW}). It should be realized that how we choose to slice up the spacetime has no effect on itself; thus in describing spacetime as a series of slices, our choice of $N(t)$ and $N^a(t)$ is completely arbitrary (except that $N(t) > 0$ always and at every point on $M$ and $N^2-N_aN^a>0$). It is in fact this choice that determines (given coordinates on $M$) what our coordinate system on spacetime will be.

We introduce $\mathcal{U}^d$ superspace - arena of dynamics for certain universe. It is an infinite dimensional manifold which contains triads $(M,g,\phi)$ as its elements.
Here $M$ is the manifold representing $d$-dimensional space, hence $g$ is well-defined Riemannian metric on $M$, and $\phi$ is the scalar field. The subset $\mathcal{U}^d_M\in\mathcal{U}^d$ i.e. all possible universes for given $d$-manifold $M$ we shall call $M$-superspace. We will denote $\mathcal{G}(\cdot,\cdot)$ the metric and $V[g,\phi]$ the potential on superspace, corresponding to the scalar field $\phi$ with with a potential $F(\phi)$.

It is important to note, that such a formulation of gravity Ref.\refcite{ADM} enables us to apply well-developed methods of the theory of dynamical systems. Our main aim is to study global behavior of cosmological models using their local, geometric properties. In particular, we are looking for equations for Hamiltonian system, which should characterize its trajectory instability, hyperbolicity, etc.

As is well known, $\mathcal{U}^d_M$ ($M$-superspace) possesses many of the usual properties of a manifold; in particular we may define the tangent and 1-form spaces of $\mathcal{U}^d_M$:
\bea 
T\mathcal{U}^d_M &\approx\, \mathcal{U}^d_M\times[S_2(M) \times S(M)]\\
T^{*}\mathcal{U}^d_M &\approx\, \mathcal{U}^d_M\times[S^2(M) \times S(M)]\notag
\eea
where $S^2(M)$ is set of 2-contravariant symmetric tensor densities, $S_2(M)$ is set of 2-covariant symmetric tensor fields and $S(M)$ - the set of scalars on $M$.

In component form a tangent vector from $T\mathcal{U}^d_M$ at the point $(g,\phi)\in\mathcal{U}^d_M$ is
written as $(k_{ab},\varphi)$. Similarly a 1-form is written as $(\pi^{ab},p)$; in this case $\pi, p$ are densities.

Having defined vectors and 1-forms on $\mathcal{U}^d_M$, the generalization to tensors of arbitrary type is straightforward, and is the same for	$\mathcal{U}^d_M$ as it is for any finite-dimensional manifold.

Thereby, we firstly define a parallel projection operator along given trajectory (smooth curve) $c$ on $\mathcal{U}^d_M$ as
\beq
 \mathcal{Q}_c=\begin{cases} 
	\dfrac{u\otimes u^\flat}{\Vert u\Vert^2}   
                 & \textrm{if }  \Vert u\Vert^2=\mathcal{G}(u,u)\neq 0 \\
	\dfrac{\partial V\otimes dV}{\Vert dV \Vert^2} 
                 & \textrm{if }  \mathcal{G}(u,u)=0
		\end{cases}
\eeq
and perpendicular projection operator as
$$
\mathcal{P}_c=I-\mathcal{Q}_c\,,
$$
where the velocity vector $u=\dot{c}=\frac{d}{dt}\big|_{c(t)}\,\in T_{c(t)}\mathcal{U}^d_M$.

In particular, the projections for given vector field $X\in T \mathcal{U}^d_M$ along $c$ are given
\beq
\mathcal{Q}_c X=\frac{\mathcal{G}(u,X)}{\mathcal{G}(u,u)}u,\qquad \mathcal{P}_c X=X-\mathcal{Q}_c X.
\eeq

Now, one can define a new derivative on superspace. A vector field $X$ is \textit{$D$-differentiable} along $c$
if there exists unique smooth vector field $D_u X$ such that if $\mathcal{G}(u,u)\ne0$ then
\beq
D_u X = \mathcal{P}_c \nabla_u \mathcal{P}_c\, X + \mathcal{Q}_c \mathcal{L}_u \mathcal{Q}_c\, X\ ,
\eeq
where $\nabla$ and $\mathcal{L}$ are the covariant and Lie derivatives, respectively.

The formula above is a natural generalization of the directional covariant and Fermi derivatives.

Then, as it is shown in Ref.\refcite{GK1} the trajectory stability of solutions of dynamical systems is reduced to the following generalized Jacobi equation
\begin{equation}
\label{eq:JacobiOmega}
D^2_u n+\Omega_u \big(n\big)=0\ ,
\end{equation}
where
\begin{equation}
\label{eq:OmegaDef}
\Omega_u\big(n\big)=R(n,u)u +\mathcal{P}_c\Bigg(\nabla_n \partial V+
\frac{3n(V)}{\mathcal{G}(u,u)}\partial V\Bigg)
\end{equation}
with initial conditions:
\beq
\label{eq:InitialCond}
\mathcal{G}(u,n)\big|_{t_0}=\mathcal{G}(u,D_un)\big|_{t_0}=0\ .
\eeq

\begin{figure}[!tb]
  \centering
    \includegraphics[width=0.8\textwidth]{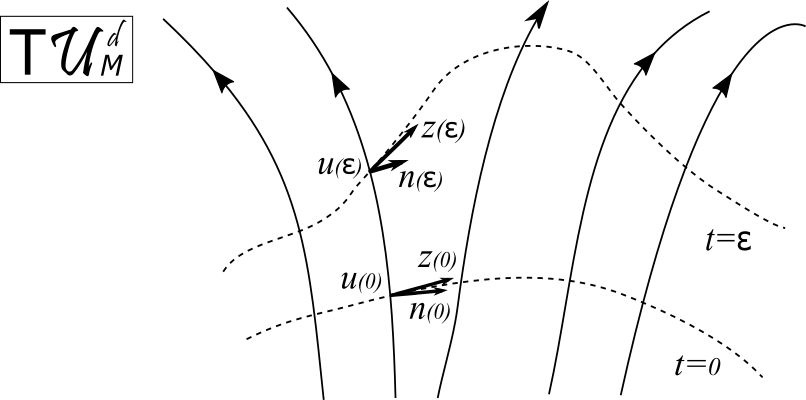}
		  \caption{The diagram illustrates the stability analysis technique.
							For a given initial perturbation $z(0)\in T\mathcal{U}^d_M$ in corresponding to $c(t)$ frame, 
							as the universe is evolved in time, $z(t)$ stretches by the motion. For a geodesic flow, 
							if perturbation is initially orthogonal to $u(0)$ in direction and velocity, then $z$ will remain orthogonal as the system evolves. However, in the case of a general flow (universe 
							with non-zero potential) for stability analysis one has to consider the ``effective'' 
							part of $z$ $n=\mathcal{P}_c z$ i.e. the perpendicular projection along trajectory.}
			\label{fig:stab}
\end{figure}

The schematic diagram in Fig.\ref{fig:stab} illustrates the described procedure. Stability analysis technique for general case of pseudo-Riemannian manifolds, and in superspace case particularly, outlined in this section has been elaborated in details in Refs \refcite{GK1,AKCMP} and the references given therein.

\section{Stability of scalar field models}
We consider minimally coupled gravitationally interacting neutral scalar with the action written as
\begin{align*}
S&=\int_W\bigg[\tfrac{1}{16\pi G} R(^4g)
-\tfrac{1}{2}g^{\mu\nu}\partial_\mu\phi\partial_\nu\phi -F(\phi)\bigg]d\mu(^4g)\\
&=\int_{I\times\Sigma}\bigg[\tfrac{1}{16\pi G}\bigg(-tr(k)^2+k\cdot k +R(g)\bigg)
+\tfrac{1}{2}\varphi^2 -\tfrac{1}{2}\|d\phi\|^2 -F(\phi)\bigg]Ndtd\mu(g)\ ,
\end{align*}
where $(g,\phi)\in\mathcal{U}^d_M,\quad(k,\varphi),\,(h,\chi)\in T_{(g,\phi)}\mathcal{U}^d_M,$ 
$d\mu(g)=(\det g)^{1/2}dx^1\wedge\dots \wedge dx^d$, 
$tr(k)=g^{ab}k_{ab}$, 
$(k\times h)_{ab}=k_{ac}g^{cd}h_{db}$, $\Vert d\phi\Vert^2=g^{ab}\phi_{|a}\phi_{|b}$.

We will introduce a metric on $\mathcal{U}^d_M$ such that the kinematical part of the Hamiltonian given by ADM formalism could be expressed by that metric Ref.\refcite{ADM}
\begin{flalign}
\mathcal{G}[g,\phi]\big(k,\varphi\,;\,h,\chi\big)&=\int_M\bigg[-tr(k)tr(h)+tr(k\times h) +\varphi\chi\bigg]d\mu(g)
\end{flalign}
and the potential
\beq
V[g,\phi]=\int_M\bigg[-\tfrac{1}{2}R(g)+\tfrac{1}{2}\big\Vert d\phi\big\Vert^2+F(\phi)\bigg]d\mu(g)
\eeq
where $8\pi G=1$, $N=1$ and $\mathbf{N}=0$ for simplicity.
The following constraint equations must hold too (cf Ref.~\refcite{AKPRD})
\begin{align}
\mathcal{H}[g,\phi;\,k,\varphi]
&=\tfrac{1}{2}\big[-tr(k)^2 +k\cdot k +\varphi^2 -R(g) +\|d\phi\|^2 +2F(\phi)\big]d\mu(g)=0,\\
\mathcal{I}[g,\phi;\, k,\varphi]
&=\big[-k^{b}{}_{a|b} +tr(k)_{|a} +\varphi\partial_a\phi\big]d\mu(g)=0\ .
\end{align}

Therefore, the perturbation vector $n$ must be tangent to the subspaces $\mathcal{H}=0$ and $\mathcal{I}=0$ of superspace; i.e. we require:
\begin{flalign}
\label{eq:H0}
(n,D_un).D_{(g,\phi,k,\varphi)}\mathcal{H}[g,\phi;\,k,\varphi]&=0\ , \\
\label{eq:I0}
(n,D_un).D_{(g,\phi,k,\varphi)}\mathcal{I}[g,\phi;\, k,\varphi]&=0\ .
\end{flalign}

If we denote $n=(h,\chi)$ and $D_un=(\tilde{h},\tilde{\chi})$, then we have
\begin{flalign}
(h,\chi,\tilde{h},\tilde{\chi}).D_{(g,\phi,k,\varphi)}
\mathcal{H}[g,\phi;\,k,\varphi]&=tr(k)tr(h\times k) 
-tr(h\times k\times k)\notag\\
&\ -\tfrac{1}{2}\big[\Delta tr(h) +\delta\delta h -tr(h\times Ric(g))\big]\notag\\
&-\tfrac{1}{2}tr(h\times d\phi\otimes d\phi) +g(d\phi,d\chi) + F'(\phi)\chi\notag\\
&\ -tr(\tilde{h})tr(k) +tr(\tilde{h}\times k) +\varphi \tilde{\chi}=0\ ,\\
(h,\chi,\tilde{h},\tilde{\chi}).D_{(g,\phi,k,\varphi)}
\mathcal{I}[g,\phi;\, k,\varphi]
&=2h^{ac}(k_{cb|a}-k_{ca|b})\notag\\
&\ +2k_{db}h^{da}{}{}_{|a}
-k^d{}_{b}tr(h)_{|d}+k^{ad}h_{ad|b}\notag \\
&\ +\varphi\chi_{|b} + \tilde{\chi}\phi_{|b} +2tr(\tilde{h})_{|b}-2\tilde{h}^a{}_{b|a}=0.
\end{flalign}

Now, $n=(h,\chi)$ is not an arbitrary perturbation, but must be perpendicular to $(k,\varphi)$ (see \eqref{eq:InitialCond}). This means that we must have:
\bea
\mathcal{G}[g,\phi]\big(k,\varphi; h,\chi\big)
=\int_M\big[-tr(k)tr(h)+tr(k\times h) +\varphi\chi\big]d\mu(g)
=0\ ,
\eea
and
\bea
\mathcal{G}[g,\phi]\big(k,\varphi; \tilde{h},\tilde{\chi}\big)
=\int_M\big[-tr(k)tr(\tilde{h})+tr(k\times \tilde{h}) +\varphi\tilde{\chi}\big]d\mu(g)
=0\ .
\eea

In order to analyse stability of the cosmological models one needs to investigate $\Omega_u \big(n\big)$ for all relevant directions and eliminate spurious perturbations. The spurious perturbations form a subspace in $T\mathcal{U}^d_M$ of a dimension equal to the number of symmetries, i.e. independent Killing 
vector fields on $M$ (see Ref.\refcite{GK1}). Therefore, after studying the spurious vectors in the appropriate 
way, only proper ones will be kept.

The purpose of the present paper does not include a derivation of the Einstein equations, as it was done in Ref.~\refcite{GK1} for FLRW universe and Ref.~\refcite{AKCMP} for general case. 



\section{Conclusions}

We defined a framework for a rigorous analysis of the global stability of cosmological solutions using a generalised Jacobi equation method of theory of dynamical systems reformulating the latter for pseudo-Riemannian manifolds by means of ADM-Hamiltonian formalism. Important feature of the used approach is the use of local geometric characteristics for the study of the stability of cosmological solutions, as distinct of e.g. Lyapunov exponents which are limiting in time quantities of the system. The derived formula are applied for the scalar field cosmological solutions in view of their basic role for the inflationary models, as well as for those of the dark energy and dark matter. More detailed treatment of the cosmological solutions using the developed approach is given in Ref. \refcite{GK1}.  

The developed formalism for the stability of solutions of Einstein equations in the superspace can be applied also to its non-cosmological solutions.


\begin{thebibliography}{0}    

\bibitem{Lin} A.D.~Linde, Particle physics and inflationary cosmology, Harwood Academic Publishers (1990).

\bibitem{Peeb} P.J.E.~Peebles, B.~Ratra, Rev. Mod. Phys. 75, 559 (2003).

\bibitem{Bar1} A.~Paliathanasis, M.~Tsamparlis, S.~Basilakos, J.D.~Barrow, Phys. Rev. D 91, 123535 (2015) 

\bibitem{Bar2} J.D.~Barrow, A.~Paliathanasis, Phys. Rev. D 94, 083518 (2016) 
   
\bibitem{Bern} T.~Bernal, V.H.~Robles, T. Matos, arXiv:1609.08644 (2016)

\bibitem{Ash} A.~Ashtekar, P.~Singh, Class. Quant. Grav. 28, 213001 (2011)
		
\bibitem{Oi} V.K.~Oikonomou,  Phys. Rev. D 94, 044004 (2016) 
    
\bibitem{MTW} C.~Misner, K.~Thorne, and J.~Wheeler: Gravitation,  W. H. Freeman (1973)

\bibitem{La} L.~Andersson, Fundam. Theor. Phys. 177, 277 (2014)

\bibitem{Ed} L.B.~Ednaldo~Junior et al, Class. Quantum Grav. 33 125006 (2016)

\bibitem{An} D.V.~Anosov, Comm. Steklov Math. Inst., 90 (1967)

\bibitem{GK1} A.V.~Gurzadyan, A.A.~Kocharyan, (work in progress) (2017)

\bibitem{AKCMP} A.A.~Kocharyan, Commun. Math. Phys. 143, 27 (1991)

\bibitem{ADM} R.~Arnowitt, S.~Deser, C. Misner, Phys. Rev. 116, 1322 (1959)

\bibitem{AKPRD} A.A.~Kocharyan, Phys. Rev. D 80, 024026 (2009)


\end{thebibliography}
\end{document}